# Collisional S-Matrix for the Vibrational Dynamics of H+$H_2$ by Quantum Computing

*Yulun Wang* and *Predrag S. Krstic*
Institute for Advanced Computational Science, Stony Brook University,
Stony Brook NY 11794-5250

**ABSTRACT**

An algorithm and a system of quantum circuits is developed and applied to compute accurately the $S$ matrix for the transitions between vibrational states of $H_2$ for collisions with H. The algorithm was applied to 100 eV laboratory collision energy at a quantum circuit simulator. The effects of the discretized dissociative continuum to the transition cross sections are carefully studied and accuracy and convergence of the results with the chosen parameters of the algorithm and the collision system are verified by comparison with a solution of the time-dependent Schrodinger equation using the classical algorithm as well as comparison with a few results from the literature.

## 1. INTRODUCTION

A Hydrogen atom colliding with a hydrogen molecule is the most fundamental neutral atom-molecule collision systems. Detailed knowledge of the inelastic and elastic processes is of high importance in the star formation processes from molecular clouds, but also in energy and particle redistribution of fusion divertor plasma, especially when hydrogen molecules are vibrationally excited. Vibrationally excited molecules have significantly enhanced dissociation. Vibrationally excited states of $H_2$ are formed by associative desorption, collisional neutralization, and excitation on the plasma-facing material surfaces, and possibly, by the three-body diatomic association in H+H+H collision. In addition, the elastic scattering at excited vibrational states of abundant hydrogen molecules can play important role in the transport and dissipation od of the divertor plasma momentum [1-6].

A typical collision event evolves through dynamically coupled electronic, vibrational, and rotational degrees of freedom. Historically, these systems have been studied thoroughly only for the processes from the ground states (electronic, vibrational, rotational) as motivated by applications, and as well as limited by experimental and theoretical capabilities of the time [3]. Also, the corresponding calculations were usually done within a manifold of bound vibrational states, thus neglecting the possible importance of inelastic processes through dissociative channels, as well as the dynamic change of the dissociative continuum edge with the relative position of the projectile and target. Importance of the effects of dissociative continuum to the transitions between the bound vibrational and electronic states were documented in [5,6].

Paul Benioff and Richard Feynman [7,8] have proposed the idea of quantum computing in the 1980s, predicting the advantages of quantum algorithms in a quantum computer for simulating the complex quantum mechanical systems. Classical computing faces difficulties in, for example, eigenvalue problems of many-body quantum systems, and transition dynamics among excited states. Despite all the challenges and difficulties to build a satisfying fault-free quantum computer in the near future, an increasing amount of effort is invested to develop quantum algorithms with this new perspective. However, the quantum computing algorithm approach for calculation of a full S matrix for the transition dynamics between bound and continuum states is still mussing.to solve the vibrational dynamics of a system in collision with the inclusion of dissociative continuum is still missing.

In this work, we aimed to build develop an algorithm and develop the system of quantum circuits for a quantum computer that would be able to calculate the evolution operator of the system, from which we obtain the $S$-matrix [5,6,8] and relevant cross-sections for the vibrational transition between dissociating vibrational states at the ground electronic potential curve of H$_2$ ($X\ ^1\Sigma_g^+ 1s\sigma$) in collision with H(1s). be able to calculate the collisional S-matrix [5,6,9] for transitions between dissociating vibrational states at the ground electronic surface of $X$ in collision with H(1s). A peculiarity of the present calculation is the inclusion of the dissociative continuum in the calculation. The finite set of the bound vibrational states can be considered as a closed quantum system in presence of the interactions with the dissociative "environment", in which case the calculated $S$ matrix will be for an open quantum system [16]. Thus, the current calculation is a possible test ground (not pursued here) for quantum computing methods for the open systems approximations in the Lindblad type of equations.

In Section 2 we define our method and prepare the algorithm to run at a quantum computer and quantum circuits for calculation of the $S$ matrix for the vibrational dynamics of H$_2$ in collision with H, including the dissociative continuum. In Section 3 we test the algorithm by simulation using quantum circuits to calculate show the S matrix and results for the relevant transition cross sections of transitions with bound vibrational states, as well as for the dissociation. These results are obtained from simulation using quantum circuits and verified by comparison with benchmark. The benchmark is obtained by solving time dependent Schrodinger equations as system of coupled ordinary differential equations (ODE), using the expansion of the H+H$_2$ wavefunction in the chosen vibrational basis of $N$ pseudo-states,

$$\Psi(\vec{r},t) = \sum_{i=1}^{N} c_i |\varphi_i\rangle \qquad (1)$$

Finally, our conclusions are presented in Section 4.

Atomic system of units, $\hbar = m_e = e = 1$, is used throughout the paper unless specified differently.

## 2. METHOD

The $H_2$ ($X\ ^1\Sigma_g^+ 1s\sigma$) together with H(1s), for various distances of the H atoms, constitutes the ground electronic surface of $H_3$ molecule. For the $H_3$ ground surface, we adopt the analytical fit of Boothroyd [10], augmented with HF-CI results of Krstic et al [1,2] at distances between any two atoms smaller than 1 a.u. The evolution operator of a quantum dynamic system is commonly approximated by trotterization method [17, 18], as explained in detail in Section 2.3. For a basis for expansion of the exponential operators we use eigenstates of the unperturbed vibrational Hamiltonian $H_0$ of the $H_2(X\ ^1\Sigma_g^+ 1s\sigma)$

$$H_0\varphi = \left(-\frac{1}{2\mu}\frac{\partial^2}{\partial r^2} + V_{H_2}(r)\right)\varphi(r,t) = \varepsilon\varphi \quad (2)$$

where $r$ is the vibrational coordinate and $\mu$ is the reduced mass of $H_2$ (~0.5 amu). This eigenvalue problem was solved with finite quantization "volume" boundary conditions, $\varphi_k^{(n)}(r_{\max}) = \varphi_k^{(n)}(r_{\min}) = 0$, where $r_{\max} = 20$ and $r_{\min} = 0.02$, with $n = 256$ mesh point along the $r$-axis. This resulted in 256 vibrational states on each of $H_2$, with 15 bound and 241 continuum (positive energy) pseudo-states. Although the relevant vibronic continua are discretized, with this choice of $r_{\max}$ the density of continuum states stays high enough even for several eV above the continuum edge. The discretization of the dissociative continuum was used earlier by Krstic et al [5,6] to calculate dissociation in the $H_3^+$ collision system at collision energies below 10 eV.

The excitation energy of the first vibrational state of $H_2$ is $\sim 0.5\ eV$, corresponding to a characteristic vibration time exceeding 50 a.u., which is comparable to the collision time of the order of $\left(\frac{\mu}{2E}\right)^{1/2}$ for energies $< 100\ eV$. The collision time is still short compared to the time scale of molecular rotations (excitation energy $0.01\ eV$, i.e., the characteristic times $>1000$ a.u.), thus enabling one to consider the direction of the diatomic internuclear axis of $H_2$ as fixed. The consequence is that the angle $\gamma$, defined as the angle between $\vec{R}$ (defined from the center of mass of the molecule to the projectile nucleus) and diatomic internuclear axis $\vec{r}$, stays constant during a collision, i.e., enters the theory as a parameter. In effect, the equations of motion are completely decupled as far as the angular variables are concerned. For the purposes of testing of our algorithm, we adopt in this work $\gamma = 90°$, as a typical representant of the atom-molecule collisional configuration, sufficient for the algorithm testing purposes. We note that for accurate calculations of an atom-diatomic molecule collision in a frozen rotation approximation, one would need to average the cross sections for various $\gamma$ in range [0, π] rad. We use Jacobi coordinates $(\vec{R}, \vec{r}, \gamma)$ for the collision geometry in this work.

The fully quantal approach, that would besides the electronic motion also quantize the motion of the hydrogen projectile is not practical at higher energies (few tens eV) because this would involve too large number of partial waves. An alternative is to replace the partial wave with the impact parameter formalism, assuming a classical motion of the projectile. The vibrational motion of $H_2$ should be still treated quantally on the lowest adiabatic electronic surface of the H₃, while H₂ rotations are frozen. The resulting time-dependent Schrodinger equation for the nuclear motion, within the Born-Oppenheimer approximation, is

$$H\Psi(R(t),r,t) = (H_0 + V(R(t),r))\Psi(R(t),r,t) = i\frac{\partial \Psi(R(t),r,t)}{\partial t} \qquad (3)$$

where $R(t)$ is the straight-line approximation, $R = \sqrt{b^2 + v^2 t^2}$, of the projectile displacement with respect to the diatomic center-of-mass, where b is the impact parameter, and v is the velocity of the projectile v=dR/dt. The v can be calculated as $v = \sqrt{\frac{2E_{CM}}{M}}$, where $E_{CM}$ is the center of mass energy, taken here to be $\frac{100}{2/3}eV$, and $M$ is the reduced mass of the H+H₂ molecule, $M \sim \frac{2}{3}$ amu. The assumption of the classical motion of the hydrogen atom at this velocity is acceptable, since its corresponding de Broglie wavelength of the hydrogen is about 0.05 a.u., much smaller than the characteristic dimension of the $H_2$ molecule and characteristic interaction radius of H and H₂. The straight-line trajectory of H and impact parameter method are acceptable approximations if the energy change ΔE in the system during the collision is much smaller than the projectile energy. Here ΔE < 4.5eV+ε_d, where ε_d is kinetic energy of dissociated atoms of H₂ (a few eV at most), and dissociation energy of the ground vibrational state of H₂ ≈ 4.5 eV. Finally,

$$V(R,r) = V_{H_3}(R,r) - V_{H_2}(r) \qquad (4)$$

where $V_{H_3}(R,r)$ is the ground potential adiabatic surface of H₃ molecule. A nice feature of $V(R,r)$ is that $V(R \to \infty, r) = 0$ which enables correct definition of the initial and final states in the S matrix.

To enable the quantum approach for solving this type of problems, Eq. 3 is firstly solved on a numerical mesh, employing the split-operator technique [19-20] in the energy representation. Evolution of the state vector in time from some large initial "$-T$":

$$\Psi(t) = e^{-i\int_{-T}^{t} H(t')dt'} \qquad (5a)$$

which can be iterated with a small timestep $\tau$, starting from $\Psi_0 = \Psi(-T)$, and following the ordering of time, as

$$\begin{aligned}\Psi_1 &= e^{-i\tau H(t_1)}\Psi_0 \\ \Psi_2 &= e^{-i\tau H(t_2)}\Psi_1 \\ &\ldots \\ \Psi_n &= e^{-i\tau H(t_n)}\Psi_{n-1}\end{aligned} \qquad (5b)$$

where $t_n = -T + n\tau$. Since $H$ is moving classically, $R = R(t)$, and $R_{max} = \sqrt{b^2 + v^2 T^2} = 20$ is chosen large enough so that $H$ is not interacting at $t = \pm T$ with $H_2$. Then $\Psi(-T) = \Psi_0$ is an unperturbed initial vibrational state, which could be any bound $\varphi_i(r)$ or a linear combination of $\varphi_i(r)$'s. In this paper we do not consider association of two atoms of H in a presence of the H projectile and dissociative state of H₂ is not used as an initial state in the calculation of the S matrix. $\tau$ is chosen small enough ($\tau = 0.01$) to have desired transition probabilities $|\langle \varphi_i | \Psi(T) \rangle|^2$ from initial $\Psi(-T)$ deviate from the benchmark ODE solution less than a given error $\varepsilon$, which will be discussed in detail in Sec. 3.

## 2.1. Preparing qubit Hamiltonian

With the discrete $N$-state vibrational basis $\{|\varphi_i\rangle\}$, defined by Eq. 2, one can write the Hamiltonian operator as:

$$H = \sum_{i,j=0}^{N-1} h_{ij} |\varphi_i\rangle\langle\varphi_j| \tag{6}$$

where the matrix elements of $H$ are defined by

$$h_{ij}(R(t)) = \langle \varphi_i | H | \varphi_j \rangle \tag{7}$$

One can use the second quantization of the Hamiltonian, which replaces the excitation operator $|\varphi_i\rangle\langle\varphi_j|$ from state $j$ to state $i$ with the corresponding operator of creation and annihilation of a population of the vibrational states, i.e., with $a_i^+ a_j$. Such procedure would require N qubits, which would require significant quantum (and also classical) computing resources when N=256. In this work, we utilize the Qubit Efficient Encoding (QEE) [11] method to map the vibrational Hamiltonian to qubit Hamiltonian. With the QEE mapping, a $N$-state system can be encoded using $n = \log_2 N = 8$ qubits, making full use of all the qubit computational basis, and establishing obvious quantum advantage of use of the computational resources. With the chosen vibrational basis set $\{|\vec{\varphi}\rangle\}$, the $i^{th}$ state of the system $|\varphi_i\rangle$ is encoded to the $i^{th}$ qubit computational basis of $n$ qubits $|q_i\rangle$, which can be expressed in individual qubits $|q_i\rangle = |y_{n-1}, \ldots, y_0\rangle$ with $y \in \{0,1\}$. Hence one can write the qubit Hamiltonian as:

$$H_q(R) = \sum_{i,j=0}^{N-1} h_{ij} |q_i\rangle\langle q_j| \tag{8}$$

$|q_i\rangle\langle q_j|$ is now the qubit excitation operator which enables the transition from state $|q_j\rangle$ to state $|q_i\rangle$: $(|q_i\rangle\langle q_j|)|q_j\rangle = |q_i\rangle$. It can be further factorized to individual qubits with the expression: $\bigotimes_{k=0}^{n-1} |y_k^i\rangle\langle y_k^j|$, where $k$ is the index of qubits, which can be replaced by Pauli operations $\{I, X, Y, Z\}$ to obtain a full qubit Hamiltonian:

$$|0\rangle\langle 0| = \frac{1}{2}(I + Z), |1\rangle\langle 1| = \frac{1}{2}(I - Z)$$
$$|0\rangle\langle 1| = \frac{1}{2}(X + iY), |1\rangle\langle 0| = \frac{1}{2}(X - iY) \tag{9}$$

After simplification, the qubit Hamiltonian can be written as a linear combination of complex coefficients with Pauli quantum operations:

$$H_q = \sum_k g_k P_k \tag{10}$$

where $g_k$ is the linear combination of $h_{pq}$ and $h_{pqrs}$ coefficients and $P_k \in \{I, X, Y, Z\}^{\otimes n}$ represent the $n$-fold tensor product of Pauli operators. The number of terms in qubit Hamiltonian $m$ follows an exponential correlation with the number of qubits $n$ used to encode the system: $m = 4^n$, which is the number of total combinations of $n$-fold Pauli operators.

For example, the qubit Hamiltonian of a general 4-state system using 2 qubit takes the form:
$$H_q = g_0 I + g_1 Z_0 + g_2 Z_1 + g_3 Z_0 Z_1 + g_4 X_0 + g_5 X_1 + g_6 Y_0 + g_7 Y_1 + g_8 X_0 Z_1$$
$$+ g_9 Y_0 Z_1 + g_{10} Z_0 X_1 + g_{11} Z_0 Y_1 + g_{12} X_0 X_1 + g_{13} Y_0 Y_1 + g_{14} X_0 Y_1 + g_{15} Y_0 X_1 \quad (11)$$

where coefficients $g_i$, $i = 0, \ldots, 15$ are time-dependent real numbers, built of $h_{ij}(t)$, and derived in Sec. S1 of SM.

## 2.2. Calculation of the matrix elements of Hamiltonian

The basic building elements of our algorithm are matrix elements of the total Hamiltonian $H$, defined in Eq. 2, using extended vibrational basis, defined by Eq. 1

$$h_{ij}(R(t)) = \langle \varphi_i | H | \varphi_j \rangle = \langle \varphi_i | H_0 + \left( V_{X^1\Sigma_g^+}(R, r) - V_{H_2}(r) \right) | \varphi_j \rangle = \varepsilon_i \delta_{ij} + V_{ij}(R) \quad (12)$$

where $\varepsilon_i$ ($i = 0, \ldots, N - 1$) are eigenenergies of the vibrational Hamiltonian $H_o$ obtained as described following Eq. 2. The lowest 15 $\varepsilon_i$'s are negaive, presenting the bound vibrational states, the other 241 are states of the discretized vibrational continuum, having positive energy. The energy spectrum of $\{\varepsilon_i\}$ is shown in Fig. 1 for the first 150 vibrational states.

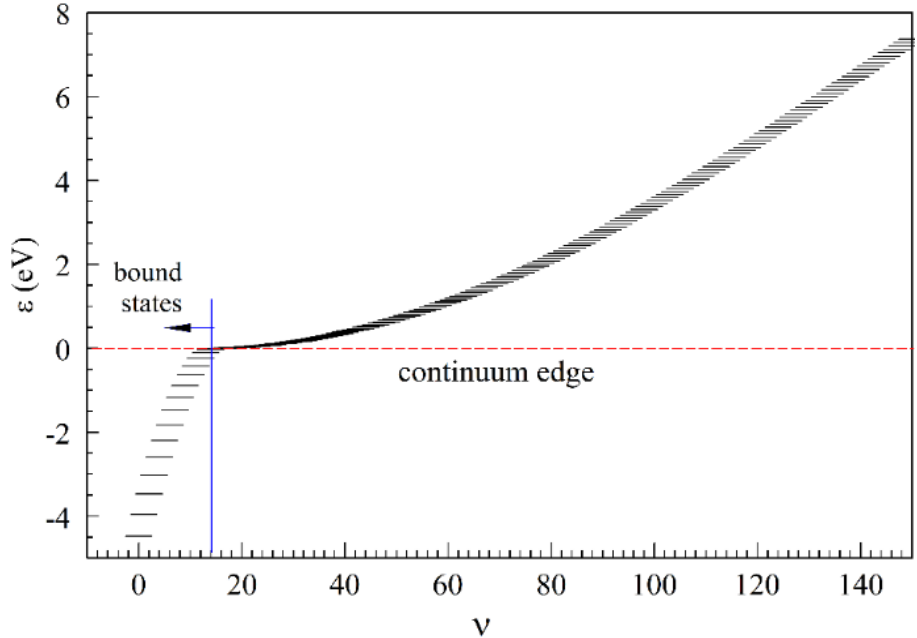

Fig. 1. Discretized energy spectrum of the first 150 vibrational states of $H_2$.

$V_{ij}(R)$ are calculated by numerical integration for all combinations $ij$, where $i = 0, \ldots, 255, j = i, \ldots, 255$, in steps $\Delta R = 0.1$, from $R = 0$ to $R = 20$. Symmetry $h_{ij}(R) = h_{ji}(R)$ is used in calculation. Several typical matrix elements (ME) $h_{ij}(R)$ are shown in Fig.2. Their values in all cases decrease fast to the negligible values when $R$ increases.

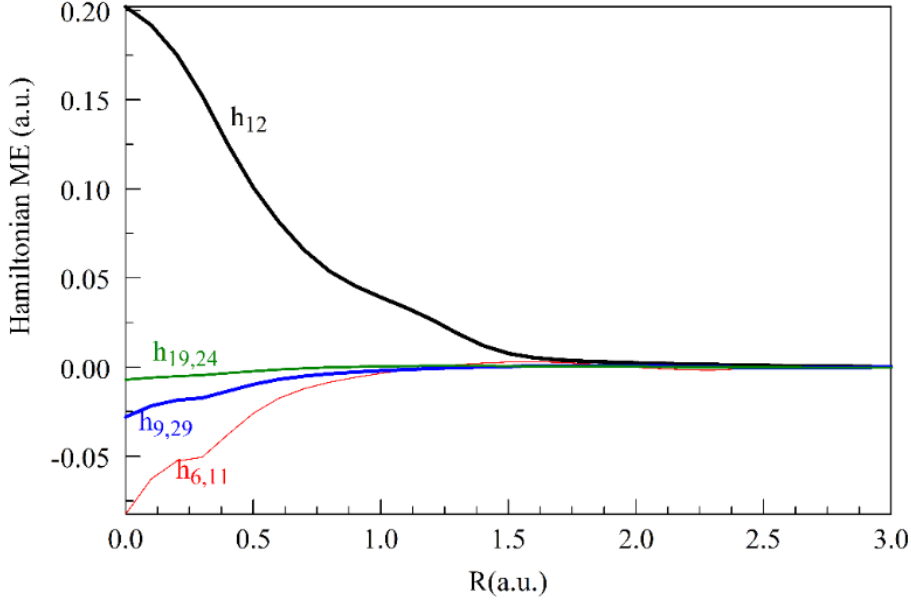

Fig. 2. Typical matrix elements $h_{ij}(R)$ of $H$ in vibrational basis.

For iteration of the system wave function by split-operator technique, the size of the time step $\tau$ was evaluated in order to find the optimal one by veryfing with the benchmark ODE solution of the time-dependent Schrodinger equation. The ODE solution has reached relative accuracy of $10^{-6}$ with time step $\tau = 0.001$. It was found that $\tau = 0.1$ has produced by split operator technique relative deviation of less than 1% from the benchamrk solution. Since the error accumulation due to the quantum noise proportional to the number of iteration steps, the application of the split operator technique can have a huge advantage when applied in the noisy quantum computer in comparison to other quantum algorithms for the time evolution of a quantum system.

In order to obtain $h_{ij}(R(t))$ at all time steps, the obtained $h_{ij}(R)$ were fitted to the desired times in the time evolution using fourth order splines. We note that the evolution time of the system was varied from $t = -T$ to $t = T$, where $T = T(b) = \sqrt{R_{\max}^2 - b^2}/v_H$ and $R_{\max} = 20$. Thus, for $b = 0.01$, we find that $V_{ij}(R_{\max}) < 10^{-6}$. The fitting to the desired time instances was done for each impact parameter, used in calculation of the cross sections.

### 2.3. Trotterization of the evolution operator

To simulate the time dependent evolution on a quantum computer, the evolution operator at time $t$, $U(t, t - \tau) = e^{-i\tau H_q(R(t))}$, needs to be applied for each timestep $\tau$, strictly following the time ordering. For simplicity, $U(t, t - \tau)$ will be written as $U(t)$ from now on. Then, following Eq. 5b, one can combine $U(t)$ at each discrete timestep together, and the overall time evolution operator $U$ is obtained as

$$U = U(T) \cdots U(-T + 2\tau)U(-T + \tau) = \prod_{t_d=T}^{-T+\tau} e^{-iH_q(R(t_d))\tau} \quad (13)$$

where $-T$ is the initial time, $T$ is final time and $t_d$'s are the discrete time points.

We note that $H_q(R_1)$ and $H_q(R_2)$ do not commute if $R_1 \neq R_2$. However, in order to calculate $U(t)$ one needs to factor it into individual Pauli operators. In general, the Pauli terms $P_k$ in a qubit Hamiltonian do not commute with each other. Hence the evolution operator $e^{-i\tau H_q(R(t))} = e^{-i\tau \sum_{k=1}^m g_k(R(t))P_k}$ cannot be exactly factored as $\prod_{k=1}^m e^{-i\tau g_k(R(t))P_k}$. A proper approximation of the evolution operator is utilized here by Trotterization, i.e., by Suzuki-Trotter approximation [21]. This approximation splits the exponential operator in a time-loaded product of simpler exponentials operators, which can be calculated for example using their eigenstates. The accuracy of the approximation depends on the small timestep $\tau$ in the evolution procedure, as well as on the number of factors used to simplify the exponential operator. Thus, for qubit Hamiltonian with $m$ terms, the second order Suzuki-Trotter gives [21]:

$$e^{-i\tau \sum_{k=1}^m g_k P_k} = \left( e^{-\frac{i\tau g_1 P_1}{2}} e^{-\frac{i\tau g_2 P_2}{2}} \cdots e^{-\frac{i\tau g_{m-1} P_{m-1}}{2}} e^{-i\tau g_m P_m} e^{-\frac{i\tau g_{m-1} P_{m-1}}{2}} \cdots e^{-\frac{i\tau g_2 P_2}{2}} e^{-\frac{i\tau g_1 P_1}{2}} \right)$$
$$+ O(\tau^3) \quad (14)$$

Each term in Eq. 14, $e^{-\frac{i\tau g_k P_k}{2}}$, can be applied by quantum circuits, discussed in detail in Subsection 2.4. Applying trotterization of a higher order can improve the accuracy, but at the cost of more terms leading to much longer computational time [12]. The choice of trotterization order should be made with careful tradeoff.

### 2.4. Quantum circuit for simulating time evolution

The operator $e^{-\frac{i\tau g_k P_k}{2}}$ in Eq. 14 with $P_k \in \{I, X, Y, Z\}^{\otimes n}$ as the $n$-fold tensor product of Pauli operators and can be applied using quantum circuits. However, the exponentiation of the $n$-fold $P_k$ is a complex quantum operation which requires a further decomposition to the elementary qubit operations such as $CNOT$ and the single qubit gates, which can be executed directly by quantum computers. The decomposition is done in two steps.

In the first step, $e^{-\frac{i\tau g_k P_k}{2}}$ is decomposed to the $n$-fold tensor product of Pauli-$Z$ and $I$ matrices $e^{-\frac{i\tau g_k A}{2}}$ ($A \in \{I, Z\}^{\otimes n}$) and a few single qubit gates [13]. For simplicity, one can express $P_k$ and $A$ in individual qubits: $P_k = \otimes \prod_{i=1}^n p_{k,i}$ and $A = \otimes \prod_{i=1}^n \alpha_i$, where $p_{k,i} \in \{I, X, Y, Z\}$ and $\alpha_i \in \{I, Z\}$ are the single qubit gates of the $i^{th}$ qubit. If $p_{k,i} = X$, $\alpha_i = Z$ and one Hadamard gate will be added on $i^{th}$ qubit on each side of $e^{-\frac{i\tau g_k A}{2}}$ operation. If $p_{k,i} = Y$, $\alpha_i = Z$ and $R_X\left(\frac{\pi}{2}\right)$ and $R_X\left(-\frac{\pi}{2}\right)$ gates will be added on $i^{th}$ qubit on the left and right side of $e^{-\frac{i\tau g_k A}{2}}$ operation respectively, where $R_X(\theta)$ is the single qubit rotation gate around $x$-axis with an angle of $\theta$ defined as:

$$R_X(\theta) = e^{-i\frac{\theta}{2}X} = I\cos\frac{\theta}{2} - iX\sin\frac{\theta}{2} = \begin{bmatrix} \cos\frac{\theta}{2} & -i\sin\frac{\theta}{2} \\ -i\sin\frac{\theta}{2} & \cos\frac{\theta}{2} \end{bmatrix} \tag{15}$$

where $I$ and $X$ are corresponded Pauli matrices. If $p_{k,i} = Z$ or $I$, $\alpha_i = Z$ or $I$ and no single qubit operation will be added on $i^{th}$ qubit. The expression of the rotational exponential operators, like in Eq. 15, in terms of a finite matrix is an advantageous consequence of the qubit encoding of the Hamiltonian.

For example, in a 4-qubit case the operator $e^{-\frac{i\tau g_k}{2}(X\otimes I\otimes Y\otimes X)}$ can be decomposed to a circuit corresponding to the operator $e^{-\frac{i\tau g_k}{2}(Z\otimes I\otimes Z\otimes Z)}$ and several single qubit gates, shown in Fig. 3.

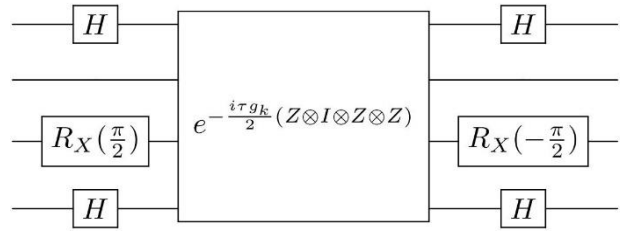

Fig. 3. An example of decomposing the operator $e^{-\frac{i\tau g_k}{2}(X\otimes I\otimes Y\otimes X)}$ to a circuit of the operator $e^{-\frac{i\tau g_k}{2}(Z\otimes I\otimes Z\otimes Z)}$ and to several single qubit gates.

In the second step, the operator $e^{-\frac{i\tau g_k A}{2}}$ is further decomposed to $CNOT$ gates and $R_Z(\tau g_k)$ rotation gates [13], where $R_Z(\theta)$ is the single qubit rotation gate around $z$-axis with an angle of $\theta$ defined as

$$R_Z(\theta) = e^{-i\frac{\theta}{2}Z} = I\cos\frac{\theta}{2} - iZ\sin\frac{\theta}{2} = \begin{bmatrix} e^{-i\frac{\theta}{2}} & 0 \\ 0 & e^{i\frac{\theta}{2}} \end{bmatrix} \tag{16}$$

This can be easily achieved by applying one $R_Z(\tau g_k)$ and connecting neighboring $i^{th}$ and $j^{th}$ qubits by $CNOT$ gate at both sides of $R_Z(\tau g_k)$ if $\alpha_i = \alpha_j = Z$. If $\alpha_i = I$, it will be omitted by $CNOT$ gates. An example of the circuit for decomposing the operator $e^{-\frac{i\tau g_k}{2}(Z\otimes I\otimes Z\otimes Z)}$ is shown in Fig. 4.

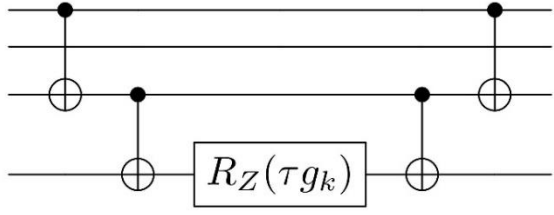

Fig. 4. An example of decomposing $e^{-\frac{i\tau g_k}{2}(Z\otimes I\otimes Z\otimes Z)}$ to a circuit of $R_Z$ gate and $CNOT$ gates.

After two steps of decomposition, the operator $e^{-\frac{i\tau g_k P_k}{2}}$ is a quantum circuit ready to be applied on a quantum computer. The evolution operator $U(t) = e^{-i\tau H_q(R(t))}$ at time $t$, which is after trotterization a product of many $e^{-\frac{i\tau g_k(R(t))P_k}{2}}$ terms, can be expressed by the quantum circuit. Furthermore, by grouping all evolution operators for each time step as a product following Eq. 13, the quantum circuit of overall evolution operator $U$ is obtained and shown in Fig. 5.

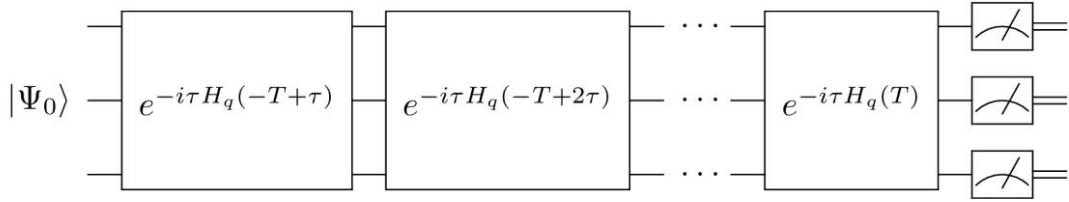

Fig. 5. The quantum circuit for computing time evolution. The circuit is initialized in state $\Psi_0$. Each box represents the evolution operator for a time step of size $\tau$. The circuit ends with the measurements of the final transition probabilities.

The evolution operator $U(t)$ shares the same structure of quantum circuits at different $t$, but different time-dependent coefficients $g_k(R(t))$ are loaded in the circuit at each time step. With an initial state $\Psi_0 = \Psi(-T)$ defined, the circuit will have the final state $\Psi(T) = U(T)\Psi(-T)$ at the end of evolution. The final transition probabilities can be obtained upon measurements of multiple shots.

With initial state defined, the probability distribution of the final states after time evolution can be obtained upon measurements, which are used in calculations in Sec. 3.

### 2.5. Calculation of the $S$-matrix

The $S$-matrix is a unitary matrix that relates the initial and final states of a physical system which undergoes a scattering process. It is important to stress that initial and final states of the system related by the S-matrix are defined in absence of the interaction which causes the transitions. Each column of the $S$-matrix contains transition amplitudes from one initial state to all possible final states.

If the initial state of the system is a mixture of populations of various eigenstates of the system, the final state will be described by the coherent combination of various $S$-matrix elements, implying the interference of the complex transition amplitudes. In this work, the S matrix element for transition from the state $|i\rangle$ to the state $|j\rangle$ is:

$$S_{ji} = \langle j|U|i\rangle = \langle j| \prod_{k=1}^{M} e^{-i\tau H_q(R(t_k))} |i\rangle \tag{17}$$

where $t_k = -T + k\tau$, $M$ is the number of time steps and $U$ is the unitary operator of the overall time evolution.

The S-matrix operator is expressed as:

$$S = \prod_{k=1}^{M} e^{-i\tau H_q(R(t_k))} \tag{18}$$

which in quantum circuit representation is equivalent to the overall time evolution operator $U$ introduced in Subsection 2.4. Following qubit efficient encoding, the $S$-matrix operator can be written as:

$$S = \sum_{i,j=0}^{N-1} S_{ij} |\mathbf{q}_i\rangle\langle \mathbf{q}_j| \tag{19}$$

where $|\mathbf{q}_i\rangle$ and $|\mathbf{q}_j\rangle$ are encoded qubit computational basis states and each element of the $S$-matrix is

$$S_{ij} = \langle \mathbf{q}_i|U|\mathbf{q}_j\rangle. \tag{20}$$

In case that initial state is a pure vibrational state, the method, shown in Section 2.4 can result to only $|S_{ij}|^2$ i.e., to probabilities from a state $|i\rangle$ to all states $|j\rangle$. In this section, we propose an original idea of a quantum module to calculate the full $S$-matrix on a quantum computer, including both phases and amplitudes, using a quantum computer, which can be considered as a variation of the Hadamard test [22]. The quantum circuit to calculate the real part of expectation value $\langle \mathbf{q}_i|U|\mathbf{q}_j\rangle$ is shown in Fig. 6. It requires two quantum registers: one register with $n$ qubits (e.g., for 256-state system, $n = 8$) enough to load the quantum states $|\mathbf{q}_i\rangle$ and $|\mathbf{q}_j\rangle$, and another register with an ancilla qubit. At the end, only ancilla qubit is measured to give the value of $Re[\langle \mathbf{q}_i|U|\mathbf{q}_j\rangle]$.

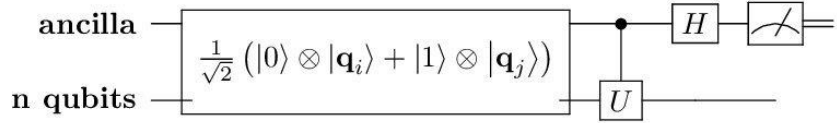

Fig. 6. The quantum circuit to calculate the $Re[\langle \mathbf{q}_i|U|\mathbf{q}_j\rangle]$. The quantum registers are initialized in the state $\frac{1}{\sqrt{2}}(|0\rangle \otimes |\mathbf{q}_i\rangle + |1\rangle \otimes |\mathbf{q}_j\rangle)$.

At the first step, the quantum circuit is initialized in the state:

$$\frac{1}{\sqrt{2}}(|0\rangle \otimes |\boldsymbol{q}_i\rangle + |1\rangle \otimes |\boldsymbol{q}_j\rangle) \tag{21}$$

The controlled-$U$ operation is applied from ancilla to other qubits, resulting in the state

$$\frac{1}{\sqrt{2}}(|0\rangle \otimes |\boldsymbol{q}_i\rangle + |1\rangle \otimes U|\boldsymbol{q}_j\rangle) \tag{22}$$

Note that $U$ is the overall evolution operator.

After a Hadamard gate applied to the ancilla qubit in Fig. 6 the quantum state becomes:

$$\frac{1}{\sqrt{2}}(|+\rangle \otimes |\boldsymbol{q}_i\rangle + |-\rangle \otimes U|\boldsymbol{q}_j\rangle)$$
$$= \frac{1}{\sqrt{2}}\left(|0\rangle \otimes (|\boldsymbol{q}_i\rangle + U|\boldsymbol{q}_j\rangle) + |1\rangle \otimes (|\boldsymbol{q}_i\rangle - U|\boldsymbol{q}_j\rangle)\right) \tag{23}$$

The probability of measuring ancilla qubit in state $|0\rangle$ is:

$$p_0 = \frac{1}{4}\left(2 + \langle \boldsymbol{q}_i|U|\boldsymbol{q}_j\rangle + \langle \boldsymbol{q}_j|U^\dagger|\boldsymbol{q}_i\rangle\right) = \frac{1}{2}\left(1 + Re[\langle \boldsymbol{q}_i|U|\boldsymbol{q}_j\rangle]\right) \tag{24}$$

Hence the value of $Re[\langle \boldsymbol{q}_i|U|\boldsymbol{q}_j\rangle]$ is equal to the difference of probabilities measuring ancilla qubit in state $|0\rangle$ and $|1\rangle$

$$Re[\langle \boldsymbol{q}_i|U|\boldsymbol{q}_j\rangle] = 2p_0 - 1 = p_0 - p_1 \tag{25}$$

Similarly, to obtain the imaginary part of expectation value $\langle \boldsymbol{q}_i|U|\boldsymbol{q}_j\rangle$: $Im[\langle \boldsymbol{q}_i|U|\boldsymbol{q}_j\rangle]$, the circuit needs to be initialized in state:

$$\frac{1}{\sqrt{2}}(|0\rangle \otimes |\boldsymbol{q}_i\rangle - i|1\rangle \otimes |\boldsymbol{q}_j\rangle) \tag{26}$$

Following the same steps as shown in Eqs. 22-25, the value of $Im[\langle \boldsymbol{q}_i|U|\boldsymbol{q}_j\rangle]$ is obtained as:

$$Im[\langle \boldsymbol{q}_i|U|\boldsymbol{q}_j\rangle] = 2p_0 - 1 = p_0 - p_1 \tag{27}$$

As an example, we show in Sec. S2 of Supplemental Material the full $S$-matrix, with both its real and imaginary parts, for bound-bound transitions, obtained for the H+H$_2$ transitions in a set of 15 bound vibrational states.

### 2.6. Computation of the ODE benchmark

By expanding the wavefunction in a set of vibrational basis states (as shown in Eq. 1), the time-dependent Schrodinger equation is approximated by a system of ODEs, which is solved by high accuracy classical methods and used as the benchmark to the results obtained by the quantum computing simulations. The set of coupled ODEs is solved using the explicit Runge-Kutta method of orders 5 and 4 (RK45, using Python *scipy.integrate.solve_ivp* function [14]). The accuracy is set by absolute and relative tolerance at each step of $10^{-12}$ and $10^{-6}$ respectively. The obtained transition amplitudes and probabilities with a time step size $\tau = 0.01\ a.u.$ are used to calculate cross-sections in Sec.3. The deviations of quantum algorithm results are compared with ODE benchmark results in the last part of Sec.3.

## 3. RESULTS

The test simulations of the algorithm in Section 2 are carried out following the quantum algorithm shown in Fig. 6 and the results are obtained by using gate-based simulator Pennylane [15].

We calculated the transition probabilities $P_{ij}$ for specific cases of initial and finals states, $|i\rangle$ and $|j\rangle$, for various values of impact parameter $b$ using quantum circuits shown in Fig. 5, as explained in Sec. 2.4. After $P_{ij}$ are obtained, the cross sections are calculated by classical integration. A cross section for transition from a bound vibrational state $i$ to a state $j$ can then be obtained as:

$$\sigma_{i-j} = 2\pi \int b P_{ij}(b) db \qquad (28)$$

We also calculate the cross sections for dissociation of a bound vibrational state $|i\rangle$ by summation of the probabilities of transition from $|i\rangle$ to all included states of dissociative continuum $|j_D\rangle$:

$$\sigma_{i-D} = 2\pi \int b \sum_{j_D} P_{ij_D}(b) \, db \qquad (29)$$

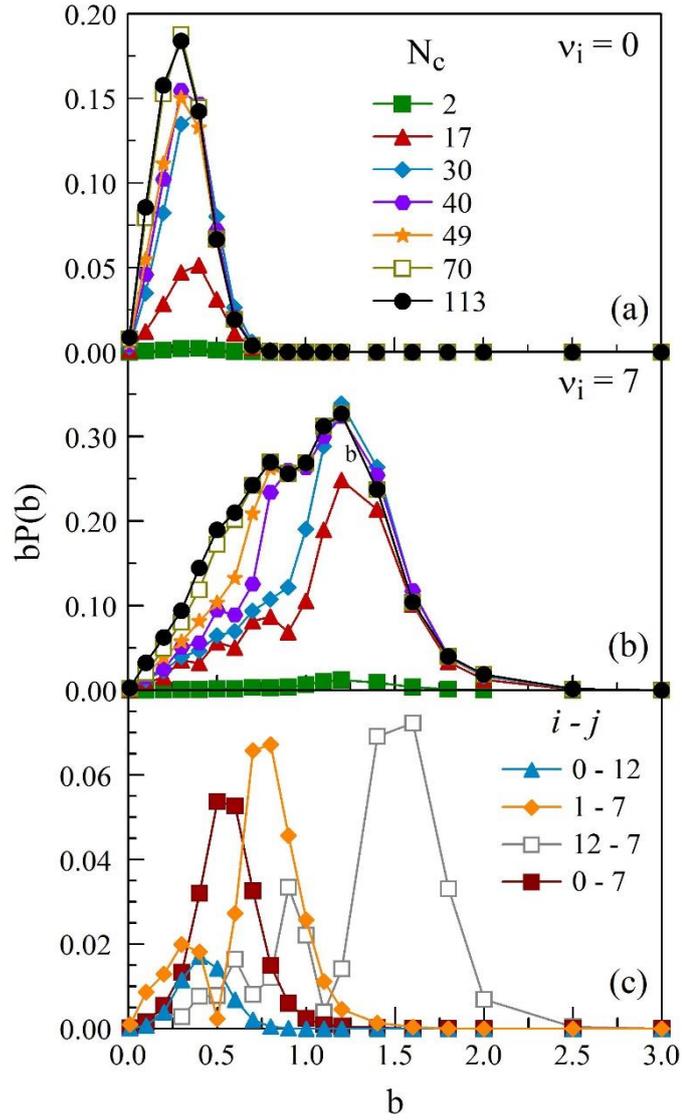

Fig. 7. $bP(b)$ as functions of $b$, for dissociation of the vibrational states (a) $i = 0$ and (b) $i = 7$, for various number of dissociative continuum states included. (c) $bP(b)$ for typical bound-bound transitions, with $N_c^{\max}$ continuum states (see text).

It is important to investigate the convergence of the bound-bound cross sections with number of included states of discrete dissociative continuum. By increasing stepwise the number of continuum states from 2, we find that this number is $N_c^{\max} = 113$. The $bP_{i-D}(b)$ for two typical cases of the initial state $|i\rangle$, $i = 0$ and $i = 7$ are shown in Fig. 7a and 7b. The $N_c^{\max}$ states span positive energy from 0 to ~5 eV. In Fig. 7c we show $bP(b)$ for a few typical bound-bound transitions, calculated with $N_c^{\max}$ continuum states. We note that unitarity of the $S$ matrix has been preserved in all calculations (to the 6 digits).

The cross sections for dissociation form various vibrational states, from the ground state ($i = 0$) to the states close to the continuum edge ($i = 12$) are shown in Fig. 8a, as functions of the number of included continuum states. The convergence of the results with 113 states, anticipated with the $bP(b)$ curves, is here confirmed at the level of cross section within the predictably uncertainty. The uncertainty is documented in Fig. 8b, which shows relative deviation (in %) of the dissociation cross sections obtained with various number of included continua from the one computed with $N_c^{max} = 113$. The deviations for $N_c = 70$ are between a fraction of ~0.5% and ~5%..

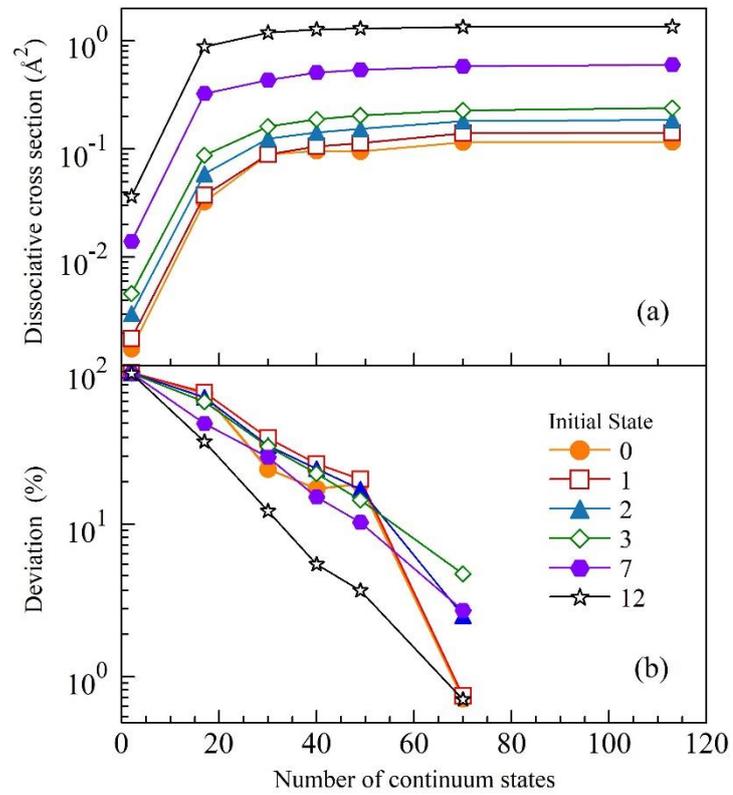

Fig. 8. (a) The cross section for dissociation versus the number of the included continuum states, from initial states $v_i = 0, 1, 2, 7, 12$. (b) The relative deviation between the cross sections of various total number of continuum states $N_c$ from the cross section with $N_c = 113$.

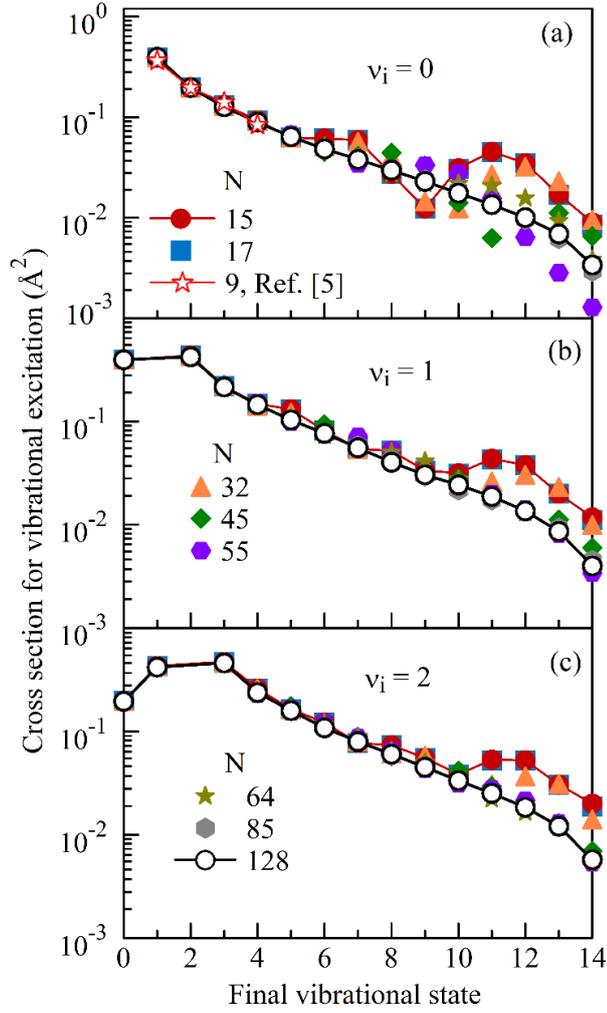

Fig. 9. The cross sections of inelastic transitions from (a) ground, (b) $1^{st}$ excited and (c) $2^{nd}$ excited vibrational state ($v_i = 0, 1, 2$) to all other vibrational bound states in presence of dissociative continuum, for various total number of continuum states $N_c = N - 15$.

The cross sections for transition from chosen initial to all bound vibrational states are shown in Figs. 9 and 10. Elastic cross sections are orders of magnitude above the inelastic cross sections shown here and are not shown. Fig. 9a also contains comparison with a fully quantum calculations of the H+H$_2$ collision system in [9], for $\gamma = 90^o$. The agreement is good. As seen in Fig. 9a, continuum states influence the transitions from the ground to the lower vibrational states only for final states larger than 5. Figs. 9b and 9c show how the couplings with the dissociative continuum influence the transitions from lower to higher vibrational states.

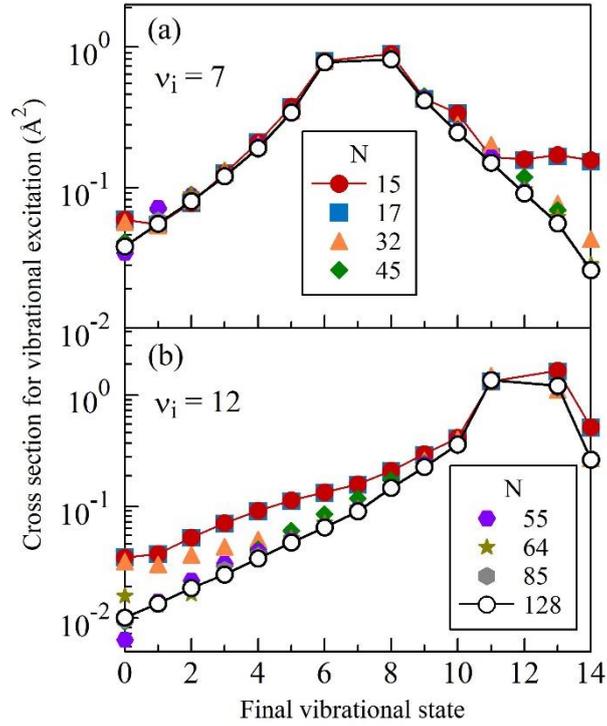

**Fig. 10**. The cross sections of inelastic transitions from (a) $7^{th}$ excited and (b) $12^{th}$ excited vibrational state ($v_i = 7, 12$) to all other vibrational bounds states in presence of dissociative continuum, for various total number of continuum states $N_c$=N-15.

These conclusions must be somewhat modified when initial state is a higher vibrational state, like are 7 and 12 in Fig. 10. For the initial state in the intermediate range ($i = 7$), significant influence of the continuum appears to the transitions to $j \geq 10$. However, if initial state is 12, transitions to both lower and higher states are persistently reduced by the presence of continuum, showing full convergence when $N_c^{\max} = 113$ continua is included in the calculation.

Fig. 11 further quantify the relative deviations of the inelastic transition cross sections among an initial and all bound vibrational states, when all $N_c^{max} = 113$ and $N_c < 113$ discretized dissociative continua are included. These deviations are close to 5% when $N_c = 70$, quantifying $N_c^{max}$ as the upper bound of the needed number of the continua for the uncertainty $< 5\ \%$.

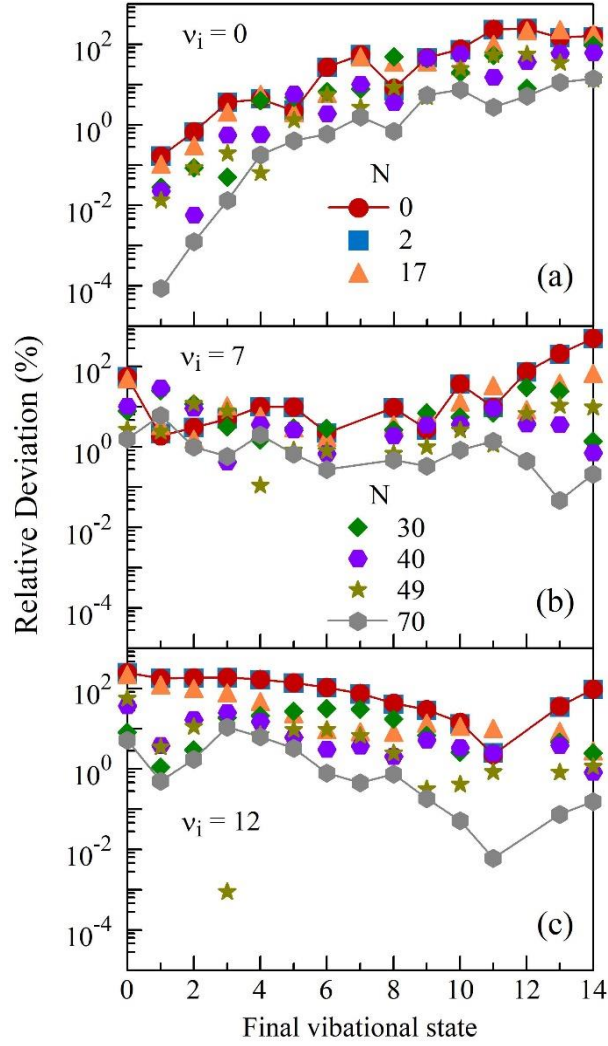

**Fig. 11**. The relative deviation between the cross sections with various number of continuum states $N_c$ included and the converged cross sections of for $N_c = 113$, for vibrational excitation from the (a) ground, (b) 7[th] and (c) 12[th] vibrational states ($v_i = 0, 7, 12$) to all other vibrational bound states.

One more factor of uncertainty deserves attention: the deviations of the results obtained using quantum circuit simulation and the benchmark results obtained by high accuracy ODE solver, as explained in Subsection 2.6. Fig. 12 shows maximal relative deviation (%, in absolute value) of the final state probabilities after evolution between these two families at $b = 0.01$. Each symbol in Fig. 12 is obtained by taking the maximal value of the relative deviations taking into account of the probabilities of all $N$ states. The datapoints are also collected from different sizes of the problem ($N = 15$ to $128$). Interestingly, we find that the main contributor to the deviation (maximal relative deviation) is the ground vibrational state, which contributes more than 5% to the uncertainty at $N = 128$. We also provided the maximal relative deviation (%) of the final state probabilities excluding the ground state data, shown in Fig. 12 as circle symbol. All other states contribute to case of calculation with 128 states at around 1.5%.

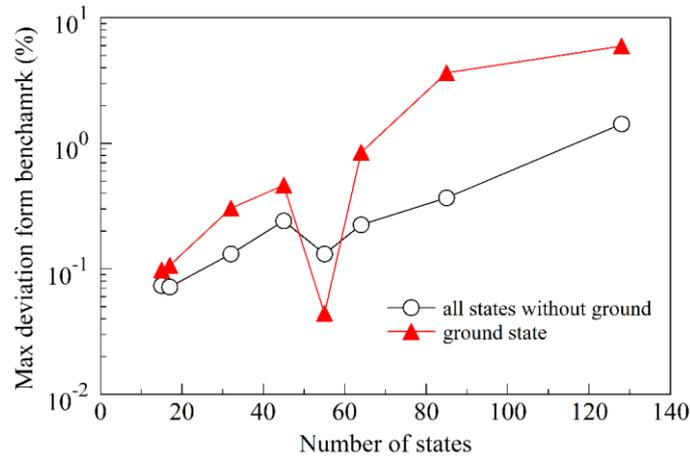

**Fig. 12**. Maximal relative deviation of the final state probabilities obtained by quantum circuit simulations from the ODE benchmark results.

**4.     CONCLUSIONS**

The algorithm for a quantum computer and corresponding quantum circuits were developed to compute the $S$ matrix, transition probabilities and cross sections for inelastic transitions within a set of bound vibrational and discretized dissociative continuum states of the ground electronic potential curve of $H_2$ colliding with H. The algorithm was tested by simulations of a fault-free quantum computer at fixed collision energy of H of 100 eV and at fixed angle of $H_2$ of 90º. Use of the Quantum Efficient Encoding to this problem of N states reduced the need for computing resources from N to $log_2$N, which also significantly reduce the errors die to the quantum noise when algorithm applied in a quantum computer. The results were verified within a few percent's deviation by comparison with the independent ODE solution of the time-dependent Schrodinger equation.

The influence on the transition vibrational dynamics of the coupling between the bound vibrational states of $H_2$ and the dissociative continuum was studied in detail within Born-Oppenheimer approximation using the proposed quantum algorithm. The convergence with the number $N_c$ of discretized continuum states at considered collision energy was reached with relative deviation of < 5% of all bound-bound as well as total dissociation cross sections with $N_c$=113, going as high and +5 eV of dissociative continuum energy. A good agreement was also reached with available literature results at the used collision energy.

# Supplemental Materials

# Collisional S-Matrix for the Vibrational Dynamics of H+H2 by Quantum Computing

*Yulun Wang* and *Predrag S. Krstic*
Institute for Advanced Computational Science, Stony Brook University,
Stony Brook NY 11794-5250

## S1. QEE encoding of qubit Hamiltonian

Taking a general 4-state system, the qubit efficient encoding (QEE) for 2 qubits and 4 states gives:

$$H_q = \sum_{i,j}^{N-1} h_{ij} |\mathbf{q}_i\rangle\langle\mathbf{q}_j|$$

$$= h_{00}|00\rangle\langle 00| + h_{11}|01\rangle\langle 01| + h_{22}|10\rangle\langle 10| + h_{33}|11\rangle\langle 11|$$
$$+ h_{01}|00\rangle\langle 01| + h_{10}|01\rangle\langle 00| + h_{12}|01\rangle\langle 10| + h_{21}|10\rangle\langle 01|$$
$$+ h_{13}|01\rangle\langle 11| + h_{31}|11\rangle\langle 01| + h_{02}|00\rangle\langle 10| + h_{20}|10\rangle\langle 00|$$
$$+ h_{03}|00\rangle\langle 11| + h_{30}|11\rangle\langle 00| + h_{23}|10\rangle\langle 11| + h_{32}|11\rangle\langle 10|$$

$$= h_{00}(|0\rangle\langle 0| \otimes |0\rangle\langle 0|) + h_{11}(|0\rangle\langle 0| \otimes |1\rangle\langle 1|) + h_{22}(|1\rangle\langle 1| \otimes |0\rangle\langle 0|) + h_{33}(|1\rangle\langle 1| \otimes |1\rangle\langle 1|)$$
$$+ h_{01}(|0\rangle\langle 0| \otimes |0\rangle\langle 1|) + h_{10}(|0\rangle\langle 0| \otimes |1\rangle\langle 0|) + h_{12}(|0\rangle\langle 1| \otimes |1\rangle\langle 0|) + h_{21}(|1\rangle\langle 0| \otimes |0\rangle\langle 1|)$$
$$+ h_{13}(|0\rangle\langle 1| \otimes |1\rangle\langle 1|) + h_{31}(|1\rangle\langle 0| \otimes |1\rangle\langle 1|) + h_{02}(|0\rangle\langle 1| \otimes |0\rangle\langle 0|) + h_{20}(|1\rangle\langle 0| \otimes |0\rangle\langle 0|)$$
$$+ h_{03}(|0\rangle\langle 1| \otimes |0\rangle\langle 1|) + h_{30}(|1\rangle\langle 0| \otimes |1\rangle\langle 0|) + h_{23}(|1\rangle\langle 1| \otimes |0\rangle\langle 1|) + h_{32}(|1\rangle\langle 1| \otimes |1\rangle\langle 0|)$$

$$= \frac{1}{4}(h_{00}+h_{11}+h_{22}+h_{33})I + \frac{1}{4}(h_{00}+h_{11}-h_{22}-h_{33})Z_0 + \frac{1}{4}(h_{00}-h_{11}+h_{22}-h_{33})Z_1$$
$$+ \frac{1}{4}(h_{00}-h_{11}-h_{22}+h_{33})Z_0 Z_1 + \frac{1}{4}(h_{13}+h_{31}+h_{02}+h_{20})X_0 + \frac{1}{4}(h_{01}+h_{10}+h_{23}+h_{32})X_1$$
$$+ \frac{i}{4}(h_{13}-h_{31}+h_{02}-h_{20})Y_0 + \frac{i}{4}(h_{01}-h_{10}+h_{23}-h_{32})Y_1 - \frac{1}{4}(h_{13}+h_{31}+h_{02}+h_{20})X_0 Z_1$$
$$- \frac{i}{4}(h_{13}-h_{31}-h_{02}+h_{20})Y_0 Z_1 - \frac{1}{4}(h_{01}+h_{10}+h_{23}+h_{32})Z_0 X_1 + \frac{i}{4}(h_{01}-h_{10}-h_{23}+h_{32})Z_0 Y_1$$
$$+ \frac{1}{4}(h_{12}+h_{21}+h_{03}+h_{30})X_0 X_1 + \frac{1}{4}(h_{12}+h_{21}-h_{03}-h_{30})Y_0 Y_1 - \frac{i}{4}(h_{12}-h_{21}-h_{03}+h_{30})X_0 Y_1$$
$$+ \frac{i}{4}(h_{12}-h_{21}+h_{03}-h_{30})Y_0 X_1$$

$$= g_0 I + g_1 Z_0 + g_2 Z_1 + g_3 Z_0 Z_1 + g_4 X_0 + g_5 X_1 + g_6 Y_0 + g_7 Y_1 + g_8 X_0 Z_1$$
$$+ g_9 Y_0 Z_1 + g_{10} Z_0 X_1 + g_{11} Z_0 Y_1 + g_{12} X_0 X_1 + g_{13} Y_0 Y_1 + g_{14} X_0 Y_1 + g_{15} Y_0 X_1 \quad (S1)$$

where

$$g_0 = \frac{1}{4}(h_{00} + h_{11} + h_{22} + h_{33}), g_1 = \frac{1}{4}(h_{00} + h_{11} - h_{22} - h_{33})$$
$$g_2 = \frac{1}{4}(h_{00} - h_{11} + h_{22} - h_{33}), g_3 = \frac{1}{4}(h_{00} - h_{11} - h_{22} + h_{33})$$
$$g_4 = \frac{1}{4}(h_{13} + h_{31} + h_{02} + h_{20}), g_5 = \frac{1}{4}(h_{01} + h_{10} + h_{23} + h_{32})$$
$$g_6 = \frac{i}{4}(h_{13} - h_{31} + h_{02} - h_{20}), g_7 = \frac{i}{4}(h_{01} - h_{10} + h_{23} - h_{32})$$
$$g_8 = -\frac{1}{4}(h_{13} + h_{31} + h_{02} + h_{20}), g_9 = -\frac{i}{4}(h_{13} - h_{31} - h_{02} + h_{20})$$
$$g_{10} = -\frac{1}{4}(h_{01} + h_{10} + h_{23} + h_{32}), g_{11} = \frac{i}{4}(h_{01} - h_{10} - h_{23} + h_{32})$$
$$g_{12} = +\frac{1}{4}(h_{12} + h_{21} + h_{03} + h_{30}), g_{13} = \frac{1}{4}(h_{12} + h_{21} - h_{03} - h_{30})$$
$$g_{14} = -\frac{i}{4}(h_{12} - h_{21} - h_{03} + h_{30}), g_{15} = \frac{i}{4}(h_{12} - h_{21} + h_{03} - h_{30}) \quad (S2)$$

## S2. S-matrix calculation by quantum algorithm

In this section, we show a full $S$-matrix, with both its real and imaginary parts, for bound-bound transitions, obtained for the H+H$_2$ transitions in a set of 15 bound vibrational states. The simulation is carried out following the quantum algorithm shown in Fig. 6 and the results are obtained from gate-based simulator, Pennylane [15].

Table S1. The S matrix elements for transitions from initial state $i = 0, 1, 2$ to final state $j = 0, 1, \ldots, 14$, obtained by quantum simulation.

|  | $i = 0$ | |  | $i = 1$ | |  | $i = 2$ | |
|---|---|---|---|---|---|---|---|---|
| $j$ | $Re[S_{ij}]$ | $Im[S_{ij}]$ | $j$ | $Re[S_{ij}]$ | $Im[S_{ij}]$ | $j$ | $Re[S_{ij}]$ | $Im[S_{ij}]$ |
| 0 | -3.199e-02 | -1.005e-01 | 0 | 1.139e-01 | -1.729e-01 | 0 | -2.632e-01 | -2.628e-02 |
| 1 | 1.142e-01 | -1.727e-01 | 1 | 3.250e-01 | -4.635e-02 | 1 | -1.595e-01 | -3.158e-01 |
| 2 | -2.631e-01 | -2.725e-02 | 2 | -1.590e-01 | -3.161e-01 | 2 | -1.424e-01 | 3.671e-01 |
| 3 | 4.297e-03 | -2.808e-01 | 3 | 2.549e-01 | -2.874e-01 | 3 | -6.341e-03 | 1.574e-01 |
| 4 | -3.332e-01 | -3.642e-03 | 4 | -1.318e-01 | -2.214e-02 | 4 | -1.826e-01 | -1.287e-01 |
| 5 | -1.917e-02 | 2.673e-01 | 5 | -1.609e-01 | -1.555e-01 | 5 | 1.059e-01 | -1.116e-01 |
| 6 | 1.345e-01 | 1.948e-01 | 6 | -9.925e-03 | 2.647e-01 | 6 | -3.887e-01 | 2.332e-01 |
| 7 | -3.636e-01 | 5.811e-02 | 7 | 9.746e-03 | 1.638e-01 | 7 | -2.664e-01 | -1.010e-01 |
| 8 | -2.574e-02 | 3.237e-01 | 8 | 1.187e-01 | -2.360e-01 | 8 | 1.366e-01 | -1.019e-01 |
| 9 | -8.424e-02 | -1.051e-01 | 9 | 1.833e-01 | 8.549e-04 | 9 | 2.828e-01 | 2.171e-01 |
| 10 | -3.778e-02 | -1.871e-01 | 10 | -1.492e-01 | -3.662e-02 | 10 | 2.000e-01 | -1.428e-01 |
| 11 | 2.399e-01 | 1.910e-01 | 11 | 6.066e-02 | -2.862e-01 | 11 | -3.038e-02 | 4.182e-02 |
| 12 | -3.007e-01 | -6.314e-03 | 12 | 2.012e-01 | 2.419e-01 | 12 | 1.014e-01 | 1.556e-02 |
| 13 | -2.076e-01 | 7.455e-02 | 13 | 2.252e-01 | 9.538e-02 | 13 | 1.200e-01 | -1.246e-02 |
| 14 | 1.540e-01 | -7.629e-02 | 14 | -1.987e-01 | -2.693e-02 | 14 | -1.344e-01 | 4.960e-02 |

Table S2. The S matrix elements for transitions from initial state $i = 3, 4, 5$ to final state $j = 0, 1, ..., 14$, obtained by quantum simulation.

|   | $i = 3$ | | | $i = 4$ | | | $i = 5$ | |
| --- | --- | --- | --- | --- | --- | --- | --- | --- |
| $j$ | $Re[S_{ij}]$ | $Im[S_{ij}]$ | $j$ | $Re[S_{ij}]$ | $Im[S_{ij}]$ | $j$ | $Re[S_{ij}]$ | $Im[S_{ij}]$ |
| 0 | 2.789e-03 | -2.809e-01 | 0 | -3.332e-01 | -1.326e-03 | 0 | -1.692e-02 | 2.674e-01 |
| 1 | 2.539e-01 | -2.882e-01 | 1 | -1.319e-01 | -2.148e-02 | 1 | -1.619e-01 | -1.545e-01 |
| 2 | -6.076e-03 | 1.574e-01 | 2 | -1.830e-01 | -1.282e-01 | 2 | 1.054e-01 | -1.121e-01 |
| 3 | 1.779e-01 | 1.138e-01 | 3 | 1.200e-01 | 2.886e-01 | 3 | 3.129e-02 | -5.670e-01 |
| 4 | 1.195e-01 | 2.888e-01 | 4 | -1.593e-01 | -5.603e-01 | 4 | 8.773e-02 | 7.962e-02 |
| 5 | 3.303e-02 | -5.669e-01 | 5 | 8.762e-02 | 7.975e-02 | 5 | 1.399e-01 | 1.756e-02 |
| 6 | -4.433e-02 | 1.508e-01 | 6 | 7.600e-03 | 4.558e-02 | 6 | -4.896e-01 | -1.516e-01 |
| 7 | 1.398e-01 | 1.493e-01 | 7 | 2.065e-01 | 2.937e-01 | 7 | -5.959e-02 | -2.491e-02 |
| 8 | 6.283e-03 | -2.801e-02 | 8 | 5.510e-02 | -2.111e-01 | 8 | 1.386e-01 | -2.118e-01 |
| 9 | -1.643e-01 | -4.441e-02 | 9 | -1.135e-01 | -1.670e-01 | 9 | 3.258e-02 | -9.773e-02 |
| 10 | -1.269e-03 | 1.167e-01 | 10 | -2.165e-01 | 1.666e-01 | 10 | -1.792e-01 | 1.181e-01 |
| 11 | -3.765e-02 | 1.351e-01 | 11 | -1.851e-02 | -8.388e-02 | 11 | -5.986e-02 | -5.851e-02 |
| 12 | -1.790e-01 | -1.510e-01 | 12 | -8.107e-02 | -3.609e-02 | 12 | -6.479e-02 | -1.059e-01 |
| 13 | -2.196e-01 | -4.277e-02 | 13 | -1.494e-01 | -5.806e-03 | 13 | -1.801e-01 | -6.441e-02 |
| 14 | 2.163e-01 | -5.176e-02 | 14 | 1.732e-01 | -9.243e-02 | 14 | 2.120e-01 | -8.092e-02 |

Table S3. The S matrix elements for transitions from initial state $i = 6, 7, 8$ to final state $j = 0, 1, ..., 14$, obtained by quantum simulation.

|   | $i = 6$ | | | $i = 7$ | | | $i = 8$ | |
| --- | --- | --- | --- | --- | --- | --- | --- | --- |
| $j$ | $Re[S_{ij}]$ | $Im[S_{ij}]$ | $j$ | $Re[S_{ij}]$ | $Im[S_{ij}]$ | $j$ | $Re[S_{ij}]$ | $Im[S_{ij}]$ |
| 0 | 1.364e-01 | 1.934e-01 | 0 | -3.630e-01 | 6.213e-02 | 0 | -2.178e-02 | 3.240e-01 |
| 1 | -7.833e-03 | 2.648e-01 | 1 | 1.125e-02 | 1.637e-01 | 1 | 1.163e-01 | -2.372e-01 |
| 2 | -3.872e-01 | 2.356e-01 | 2 | -2.671e-01 | -9.899e-02 | 2 | 1.357e-01 | -1.031e-01 |
| 3 | -4.366e-02 | 1.510e-01 | 3 | 1.406e-01 | 1.485e-01 | 3 | 6.091e-03 | -2.805e-02 |
| 4 | 7.730e-03 | 4.556e-02 | 4 | 2.077e-01 | 2.928e-01 | 4 | 5.398e-02 | -2.114e-01 |
| 5 | -4.898e-01 | -1.509e-01 | 5 | -5.966e-02 | -2.475e-02 | 5 | 1.378e-01 | -2.123e-01 |
| 6 | -1.409e-02 | -3.129e-02 | 6 | -2.091e-01 | -1.798e-01 | 6 | 3.014e-01 | 6.866e-02 |
| 7 | -2.089e-01 | -1.801e-01 | 7 | -3.194e-01 | 7.004e-02 | 7 | -2.550e-01 | -1.170e-01 |
| 8 | 3.012e-01 | 6.939e-02 | 8 | -2.549e-01 | -1.173e-01 | 8 | 3.084e-01 | 2.475e-01 |
| 9 | 2.925e-02 | 1.614e-02 | 9 | -1.241e-01 | -3.961e-01 | 9 | 3.189e-03 | -2.184e-01 |
| 10 | -2.688e-01 | 1.062e-01 | 10 | 2.371e-02 | 5.973e-02 | 10 | -5.059e-03 | 3.494e-01 |
| 11 | -4.963e-02 | -2.098e-01 | 11 | 4.170e-02 | 1.117e-01 | 11 | -2.737e-01 | 1.004e-01 |
| 12 | 3.847e-02 | -5.121e-02 | 12 | -6.698e-02 | 1.180e-01 | 12 | -9.084e-03 | -5.146e-02 |
| 13 | -1.231e-01 | -9.098e-02 | 13 | 1.196e-01 | 1.893e-01 | 13 | 7.443e-02 | 1.520e-01 |
| 14 | 1.998e-01 | -8.642e-02 | 14 | -2.435e-01 | 3.607e-02 | 14 | -2.530e-01 | 4.975e-02 |

Table S4. The S matrix elements for transitions from initial state $i = 9, 10, 11$ to final state $j = 0, 1, ..., 14$, obtained by quantum simulation.

| | $i = 9$ | | | $i = 10$ | | | $i = 11$ | |
|---|---|---|---|---|---|---|---|---|
| $j$ | $Re[S_{ij}]$ | $Im[S_{ij}]$ | $j$ | $Re[S_{ij}]$ | $Im[S_{ij}]$ | $j$ | $Re[S_{ij}]$ | $Im[S_{ij}]$ |
| 0 | -8.563e-02 | -1.040e-01 | 0 | -4.043e-02 | -1.866e-01 | 0 | 2.427e-01 | 1.874e-01 |
| 1 | 1.833e-01 | -1.230e-03 | 1 | -1.497e-01 | -3.478e-02 | 1 | 5.691e-02 | -2.869e-01 |
| 2 | 2.848e-01 | 2.144e-01 | 2 | 1.985e-01 | -1.449e-01 | 2 | -2.990e-02 | 4.216e-02 |
| 3 | -1.646e-01 | -4.312e-02 | 3 | -2.391e-04 | 1.167e-01 | 3 | -3.635e-02 | 1.355e-01 |
| 4 | -1.146e-01 | -1.663e-01 | 4 | -2.153e-01 | 1.682e-01 | 4 | -1.919e-02 | -8.373e-02 |
| 5 | 3.211e-02 | -9.789e-02 | 5 | -1.785e-01 | 1.191e-01 | 5 | -6.024e-02 | -5.811e-02 |
| 6 | 2.930e-02 | 1.604e-02 | 6 | -2.684e-01 | 1.074e-01 | 6 | -5.072e-02 | -2.095e-01 |
| 7 | -1.250e-01 | -3.958e-01 | 7 | 2.389e-02 | 5.965e-02 | 7 | 4.214e-02 | 1.115e-01 |
| 8 | 2.961e-03 | -2.184e-01 | 8 | -4.371e-03 | 3.495e-01 | 8 | -2.734e-01 | 1.011e-01 |
| 9 | -3.233e-01 | -3.718e-01 | 9 | 6.286e-02 | 1.553e-02 | 9 | 3.286e-01 | -1.807e-01 |
| 10 | 6.285e-02 | 1.559e-02 | 10 | -1.344e-01 | 5.638e-01 | 10 | 1.188e-01 | 1.651e-01 |
| 11 | 3.289e-01 | -1.802e-01 | 11 | 1.187e-01 | 1.652e-01 | 11 | 5.130e-01 | -2.120e-01 |
| 12 | 1.147e-01 | 2.298e-01 | 12 | 1.110e-01 | 1.887e-01 | 12 | 1.527e-01 | -2.267e-01 |
| 13 | 1.113e-01 | -1.446e-01 | 13 | 2.313e-01 | -1.687e-01 | 13 | -1.905e-01 | 8.262e-02 |
| 14 | 1.812e-01 | -9.086e-03 | 14 | 1.365e-01 | 7.377e-02 | 14 | -1.383e-01 | -1.049e-01 |

Table S5. The S matrix elements for transitions from initial state $i = 12, 13, 14$ to final state $j = 0, 1, ..., 14$, obtained by quantum simulation.

| | $i = 12$ | | | $i = 13$ | | | $i = 14$ | |
|---|---|---|---|---|---|---|---|---|
| $j$ | $Re[S_{ij}]$ | $Im[S_{ij}]$ | $j$ | $Re[S_{ij}]$ | $Im[S_{ij}]$ | $j$ | $Re[S_{ij}]$ | $Im[S_{ij}]$ |
| 0 | -3.008e-01 | -1.613e-03 | 0 | -2.064e-01 | 7.788e-02 | 0 | 1.527e-01 | -7.880e-02 |
| 1 | 2.045e-01 | 2.391e-01 | 1 | 2.265e-01 | 9.217e-02 | 1 | -1.991e-01 | -2.404e-02 |
| 2 | 1.016e-01 | 1.434e-02 | 2 | 1.199e-01 | -1.395e-02 | 2 | -1.337e-01 | 5.130e-02 |
| 3 | -1.806e-01 | -1.492e-01 | 3 | -2.200e-01 | -4.041e-02 | 3 | 2.157e-01 | -5.414e-02 |
| 4 | -8.138e-02 | -3.539e-02 | 4 | -1.494e-01 | -4.437e-03 | 4 | 1.723e-01 | -9.406e-02 |
| 5 | -6.555e-02 | -1.054e-01 | 5 | -1.806e-01 | -6.302e-02 | 5 | 2.114e-01 | -8.261e-02 |
| 6 | 3.817e-02 | -5.144e-02 | 6 | -1.237e-01 | -9.020e-02 | 6 | 1.993e-01 | -8.774e-02 |
| 7 | -6.644e-02 | 1.183e-01 | 7 | 1.205e-01 | 1.886e-01 | 7 | -2.433e-01 | 3.736e-02 |
| 8 | -9.259e-03 | -5.143e-02 | 8 | 7.502e-02 | 1.517e-01 | 8 | -2.527e-01 | 5.080e-02 |
| 9 | 1.152e-01 | 2.295e-01 | 9 | 1.109e-01 | -1.449e-01 | 9 | 1.812e-01 | -9.652e-03 |
| 10 | 1.112e-01 | 1.886e-01 | 10 | 2.310e-01 | -1.691e-01 | 10 | 1.366e-01 | 7.347e-02 |
| 11 | 1.525e-01 | -2.268e-01 | 11 | -1.904e-01 | 8.284e-02 | 11 | -1.385e-01 | -1.047e-01 |
| 12 | -4.509e-01 | -4.222e-01 | 12 | 3.428e-01 | -3.140e-03 | 12 | 2.719e-02 | 9.022e-02 |
| 13 | 3.428e-01 | -2.976e-03 | 13 | -2.681e-02 | 4.823e-01 | 13 | 3.942e-01 | 3.422e-02 |
| 14 | 2.712e-02 | 9.024e-02 | 14 | 3.942e-01 | 3.433e-02 | 14 | 4.903e-01 | -3.207e-01 |